**Comment on "Control of Spin Precession in a Spin-Injected Field Effect Transistor"**

**[Science, 325, 1515 (2009)]**

Supriyo Bandyopadhyay[*]

Department of Electrical and Computer Engineering

Virginia Commonwealth University

Richmond, VA 23284, USA

## ABSTRACT

A recently published report in the journal *Science* claims that the Datta-Das Spin Field Effect Transistor has been demonstrated because an *exact* agreement was found between the voltage modulation in a fabricated structure and a theoretical equation that supposedly describes the voltage modulation of the Datta-Das Transistor. Here, I show that the said theoretical equation holds only for a one-dimensional transistor, or perhaps a quasi one-dimensional transistor, and is certainly *not* applicable to the two-dimensional transistor studied. Hence, the exact agreement is meaningless and does not substantiate the claim.

[*] Corresponding author. E-mail: sbandy@vcu.edu

In a recently published report in *Science* (*1*), the authors claim that they have demonstrated the Datta-Das Spin Field Effect Transistor (*2*) in an electrically gated *two-dimensional* electron gas with ferromagnetic source and drain contacts. The claim is based on the observation of a non-local voltage modulation $\Delta V$ at the drain contact that exhibits an oscillatory dependence on the gate controlled Rashba interaction strength $\alpha$ and can be fitted *exactly* with the equation

$$\Delta V = A\cos\left(2m^{*}\alpha L/\hbar^{2} + \varphi\right) \qquad (1)$$

where the symbols have the same meaning as in (*1*). Reference (*1*) cites (*2*) as the source for Equation (1) [see their supporting material] and asserts that the theory of (*2*) leads to this equation for the voltage modulation in the Datta-Das transistor. Accordingly, exact agreement between Equation (1) and the observed non-local voltage modulation is offered as proof that the Datta-Das Transistor has been demonstrated.

The Datta-Das paper (*2*) however does not contain Equation (1) and this equation is actually incorrect for the two-dimensional transistor channel studied in (*1*). It is only correct for a strictly one dimensional channel and perhaps approximately correct for a quasi one-dimensional channel whose width is much smaller than $\hbar^{2}/(m^{*}\alpha)$ [ $m^{*}$ is the carrier's effective mass]. A careful reading of the Datta-Das paper (*2*) should have made this obvious. Reference (*2*) never derived any equation like (1) for a two-dimensional channel, but the correct equation for such a system, assuming perfect spin injection and detection at the source and drain contacts, will be [see Appendix 1]

$$\Delta V = \iint\int dk_{xF}\,dk_{zF}\,d\left(\Delta k_{zF}\right) B\left(k_{xF},k_{zF}\right)\cos\left[\theta\left(k_{xF},k_{zF},\Delta k_{zF},\alpha\right)L\right], \qquad (2)$$

where $k_{xF}$ and $k_{zF}$ denote Fermi wavevector components in the plane of the two-dimensional channel (assumed to be the x-z plane with charge current flowing in the x-direction), $\Delta k_{zF}$ is the difference between the z-components of the Fermi wavevectors in two spin-split subbands in the presence of Rashba interaction, $B(k_{xF}, k_{zF})$ is a function of $k_{xF}$ and $k_{zF}$, and $\theta(k_{xF}, k_{zF}, \Delta k_{zF}, \alpha)L$ is the phase shift between the orthogonal spin states in the transistor channel. It is a function of $k_{xF}, k_{zF}, \Delta k_{zF}$ and $\alpha$.

The triple integration in Equation (2) represents ensemble averaging over the two transverse Fermi wavevector components and $\Delta k_{zF}$ in a two-dimensional system. This averaging happens even at a temperature of 0 K and in ballistic transport. It inevitably dilutes the voltage modulation of the two-dimensional Datta-Das transistor because $\theta(k_{xF}, k_{zF}, \Delta k_{zF}, \alpha)$ is a function of $k_{xF}, k_{zF}, \Delta k_{zF}$. This point was already made in (*2*), although not as explicitly as it is made here. Ref. (*2*) then showed that if the channel width is much less than $\hbar^2/(m^*\alpha)$, then the phase shift $\theta L$ will be approximately wavevector-independent and given by $2m^*\alpha L/\hbar^2$, so that only in such a narrow quasi one-dimensional channel, Equation (1) can approximately apply. If the channel width is not much narrower than $\hbar^2/(m^*\alpha)$, then Equation (1) *cannot* apply.

Based on the reported values of $m^*$ and $\alpha$ in (*1*), $\hbar^2/(m^*\alpha) \leq 0.188$ μm, whereas the device width in (*1*) was 8 μm. Therefore, the devices in (*1*) were clearly two-dimensional where Equation (1) cannot apply, but Equation (2) will.

Note that Equation (2) can never be recast as Equation (1) because not only is $\theta(k_{xF}, k_{zF}, \Delta k_{zF}, \alpha) \neq 2m^*\alpha L/\hbar^2 + \varphi$, but also $\theta(k_{xF}, k_{zF}, \Delta k_{zF}, \alpha)$ is a function of $k_{xF}, k_{zF}, \Delta k_{zF}$ so that the cosine term in Equation (2) cannot be pulled outside the integral. Therefore, Equations (1) and (2) are entirely different. As a result, the voltage modulation expected in a <u>two-dimensional channel</u> is entirely different from Equation (1) and will not exhibit a cosine modulation since the triple integration will modify the cosine function.

Since (*1*) has based its entire claim on the precise agreement between the experimentally observed $\Delta V$ and Equation (1), as opposed to the correct Equation (2), we must conclude that the claim of having demonstrated the Datta-Das transistor is not sustainable.

**References**

1. H. C. Koo, J. H. Kwon, J. Eom, J. Chang, S. H. Han and M. Johnson, *Science*, **325**, 1515 (2009)
2. S. Datta and B. Das, *Appl. Phys. Lett.*, **56**, 665 (1990).
3. S. Bandyopadhyay and M. Cahay, *Introduction to Spintronics*, (CRC Press, Boca Raton, 2008).

**Appendix 1: Derivation of Equation (2)**

Consider the two-dimensional channel of a Spin Field Effect Transistor (SPINFET) in the x-z plane [†]. The wavevectors in the x- and z-directions are designated as $k_x$ and $k_z$, and $k_t^2 = k_x^2 + k_z^2$ as shown below.

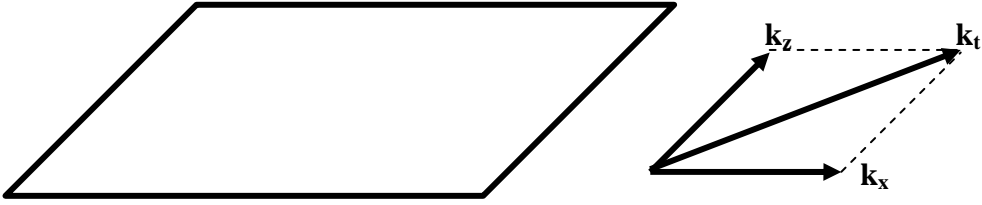

Fig. 1: Wavevector components in two dimensions.

We start with the Hamiltonian for the two-dimensional channel of the SPINFET given in ref. (*2*) where the gate field inducing the Rashba spin splitting is in the y-direction and the channel is in the x-z plane:

$$H = \begin{bmatrix} \dfrac{\hbar^2 k_t^2}{2m^*} + \alpha k_x & -\alpha k_z \\ -\alpha k_z & \dfrac{\hbar^2 k_t^2}{2m^*} - \alpha k_x \end{bmatrix}. \qquad (A1)$$

---

[†] Ref. (*1*) called the two-dimensional plane the x-y plane, whereas we call it the x-z plane. This makes no physical difference whatsoever since the axis designation is completely arbitrary. Our choice conforms to the choice of Ref. (*2*) and allows us to use the Hamiltonian from ref. (*2*) directly.

Diagonalization of this Hamiltonian yields the eigenenergies and eigenspinors in the two spin-split bands:

$$E = \frac{\hbar^2 k_t^2}{2m^*} - \alpha k_t \text{ (lower band)}; \quad E = \frac{\hbar^2 k_t^2}{2m^*} + \alpha k_t \text{ (upper band)}$$

$$[\Psi] = \begin{bmatrix} \sin\phi \\ \cos\phi \end{bmatrix} \text{ (lower band)}; \quad [\Psi] = \begin{bmatrix} -\cos\phi \\ \sin\phi \end{bmatrix} \text{ (upper band)} \tag{A2}$$

where $\phi = \frac{1}{2}\arctan\left(\frac{k_z}{k_x}\right)$. These expressions agree with those given in ref. (3).

In the SPINFET configuration, the source contact is polarized in the x-direction and injects +x-polarized spins. For the sake of simplicity, we will assume that the spin injection efficiency is 100%, i.e. only +x-polarized spins are injected into the channel at the complete exclusion of –x-polarized spins. The injected spin will couple into the two spin eigenstates in the channel. It is as if the x-polarized beam splits into two beams, each corresponding to one of the channel eigenspinors:

$$\underbrace{\frac{1}{\sqrt{2}}\begin{bmatrix} 1 \\ 1 \end{bmatrix}}_{\text{source spinor}} = C_1 \begin{bmatrix} \sin\phi \\ \cos\phi \end{bmatrix} + C_2 \begin{bmatrix} -\cos\phi \\ \sin\phi \end{bmatrix} \tag{A3}$$

where the coupling coefficients are found by solving Equation (A3). The result is

$$C_1 = C_1(k_x, k_z) = \sin\left(\phi + \frac{\pi}{4}\right)$$
$$C_2 = C_2(k_x, k_z) = -\cos\left(\phi + \frac{\pi}{4}\right) \tag{A4}$$

Note that the coupling coefficients depend on $k_x$ and $k_z$.

At the drain end, the two beams interfere to yield the spinor of the electron impinging on the drain:

$$[\psi]_{drain} = C_1 \begin{bmatrix} \sin\phi \\ \cos\phi \end{bmatrix} e^{ik_x^{(1)}L} + C_2 \begin{bmatrix} -\cos\phi \\ \sin\phi \end{bmatrix} e^{ik_x^{(2)}L}$$

$$= \sin\left(\phi+\frac{\pi}{4}\right)\begin{bmatrix} \sin\phi \\ \cos\phi \end{bmatrix} e^{ik_x^{(1)}L} - \cos\left(\phi+\frac{\pi}{4}\right)\begin{bmatrix} -\cos\phi \\ \sin\phi \end{bmatrix} e^{ik_x^{(2)}L} \quad (A5)$$

$$= \begin{bmatrix} \sin\left(\phi+\frac{\pi}{4}\right)\sin\phi\, e^{ik_x^{(1)}L} + \cos\left(\phi+\frac{\pi}{4}\right)\cos\phi\, e^{ik_x^{(2)}L} \\ \sin\left(\phi+\frac{\pi}{4}\right)\cos\phi\, e^{ik_x^{(1)}L} - \cos\left(\phi+\frac{\pi}{4}\right)\sin\phi\, e^{ik_x^{(2)}L} \end{bmatrix}$$

where $k_x^{(1)}$ and $k_x^{(2)}$ are the x-components of the wavevectors of our electron in the two spin-split bands and $L$ is the channel length (distance between source and drain contacts).

Since the drain is polarized in the same orientation as the source, it transmits only +x-polarized spins, so that spin filtering at the drain will yield a non-local voltage $V \propto |P|^2$, where $P$ is the projection of the impinging spinor on the eigenspinor of the drain. It is given by

$$P = \underbrace{\frac{1}{\sqrt{2}}[1\ 1]}_{\text{Hermitian conjugate of drain spinor}} \begin{bmatrix} \sin\left(\phi+\frac{\pi}{4}\right)\sin\phi\, e^{ik_x^{(1)}L} + \cos\left(\phi+\frac{\pi}{4}\right)\cos\phi\, e^{ik_x^{(2)}L} \\ \sin\left(\phi+\frac{\pi}{4}\right)\cos\phi\, e^{ik_x^{(1)}L} - \cos\left(\phi+\frac{\pi}{4}\right)\sin\phi\, e^{ik_x^{(2)}L} \end{bmatrix}$$

$$= \frac{1}{\sqrt{2}}\left\{\sin\left(\phi+\frac{\pi}{4}\right)[\sin\phi+\cos\phi]e^{ik_x^{(1)}L} + \cos\left(\phi+\frac{\pi}{4}\right)[\cos\phi-\sin\phi]e^{ik_x^{(2)}L}\right\} \quad (A6)$$

$$= \sin^2\left(\phi+\frac{\pi}{4}\right)e^{ik_x^{(1)}L} + \cos^2\left(\phi+\frac{\pi}{4}\right)e^{ik_x^{(2)}L}$$

Here, we have assumed 100% spin filtering efficiency.

Therefore,

$$V \propto \cos^4\left(\phi+\frac{\pi}{4}\right)\left|1+\tan^2\left(\phi+\frac{\pi}{4}\right)e^{i\left[k_x^{(1)}-k_x^{(2)}\right]L}\right|^2 = \cos^4\left(\phi+\frac{\pi}{4}\right)\left|1+\tan^2\left(\phi+\frac{\pi}{4}\right)e^{i\theta L}\right|^2$$

$$\text{or, } V \propto \cos^4\left(\phi+\frac{\pi}{4}\right)\left[1+\tan^4\left(\phi+\frac{\pi}{4}\right)+2\tan^2\left(\phi+\frac{\pi}{4}\right)\cos(\theta L)\right] \quad (A7)$$

$$\text{or, } V \propto \left[\cos^4\left(\phi+\frac{\pi}{4}\right)+\sin^4\left(\phi+\frac{\pi}{4}\right)+\frac{1}{2}\cos^2(2\phi)\cos(\theta L)\right]$$

where $\theta = k_x^{(1)} - k_x^{(2)} = \Delta k_x$.

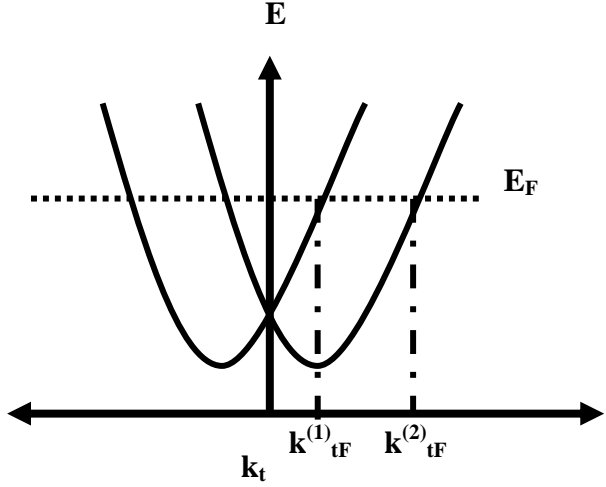

Fig. 2: Dispersion relations of Rashba spin split bands in two dimensions.

Since the experiment in (*1*) is carried out at low temperatures and presumably under low electrical bias, spin transport occurs in the Fermi circle. Therefore, we have to evaluate $\theta$ at the Fermi energy. The dispersion relations in Equation (A2) are plotted above. From these relations, we find that the Fermi wavevectors in the upper and lower bands (which are actually horizontally displaced from each other along the wavevector axis) are related according to $k_{tF}^{(1)} - k_{tF}^{(2)} = \pm 2m^*\alpha/\hbar^2$. Expressing the Fermi wavevectors in terms of their x- and z-components, we get:

$$\sqrt{\left[k_{xF}^{(1)}\right]^2 + \left[k_{zF}^{(1)}\right]^2} - \sqrt{\left[k_{xF}^{(2)}\right]^2 + \left[k_{zF}^{(2)}\right]^2} = \pm 2m^*\alpha/\hbar^2. \quad (A8)$$

Defining $k_{xF} = \dfrac{k_{xF}^{(1)} + k_{xF}^{(2)}}{2}$, $k_{zF} = \dfrac{k_{zF}^{(1)} + k_{zF}^{(2)}}{2}$ and $\Delta k_{zF} = k_{zF}^{(1)} - k_{zF}^{(2)}$, we see from the above equation that at the Fermi energy, $\theta(=\Delta k_{xF})$ is a function of $k_{xF}, k_{zF}, \Delta k_{zF}$ and $\alpha$. Equation (A7) should therefore be written explicitly as

$$V(k_{xF},k_{zF},\Delta k_{zF},\alpha) \propto \left[\cos^4\left(\phi(k_{xF},k_{zF})+\frac{\pi}{4}\right) + \sin^4\left(\phi(k_{xF},k_{zF})+\frac{\pi}{4}\right) + \frac{1}{2}\cos^2(2\phi(k_{xF},k_{zF}))\cos\{\theta(k_{xF},k_{zF},\Delta k_{zF},\alpha)L\}\right]$$
(A9)

The average value of the non-local potential that will be measured in an experiment as a function of $\alpha$ is obtained by averaging $V(k_{xF},k_{zF},\Delta k_{zF},\alpha)$ over $k_{xF}, k_{zF}$ and $\Delta k_{zF}$ on the Fermi circle, so that

$$\langle V\rangle(\alpha) \propto \left\langle \cos^4\left(\phi(k_{xF},k_{zF})+\frac{\pi}{4}\right)\right\rangle + \left\langle \sin^4\left(\phi(k_{xF},k_{zF})+\frac{\pi}{4}\right)\right\rangle$$
$$+ \left\langle \frac{1}{2}\cos^2(2\phi(k_{xF},k_{zF}))\cos\{\theta(k_{xF},k_{zF},\Delta k_{zF},\alpha)L\}\right\rangle \quad (A10)$$

or, $\langle V\rangle(\alpha) \propto \left\langle \dfrac{1}{2}\cos^2(2\phi(k_{xF},k_{zF}))\cos\{\theta(k_{xF},k_{zF},\Delta k_{zF},\alpha)L\}\right\rangle +$ constant independent of $\alpha$.

where the angular brackets denote ensemble averaging over $k_{xF}, k_{zF}, \Delta k_{zF}$ on the Fermi circle. Therefore, the $\alpha$-dependent modulating voltage measured in the experiment of (*1*) should be expressed as

$$\Delta V\big|_{measured} = \iiint dk_{xF}\, dk_{zF}\, d(\Delta k_{zF})\, B(k_{xF},k_{zF})\cos[\theta(k_{xF},k_{zF},\Delta k_{zF},\alpha)L]. \quad (A11)$$

where the ensemble averaging is effected via the integration over the variables on the Fermi circle. Equation (A11) is Equation (2) in the Comment.